\documentclass[twocolumn,showpacs,preprintnumbers,amsmath,amssymb,prx,superscriptaddress,longbibliography]{revtex4-1}
\usepackage{bbm}
\usepackage{mathrsfs}
\usepackage{graphicx}
\usepackage{dcolumn}
\usepackage{bm}
\usepackage{braket}
\usepackage{amsmath}
\usepackage{amsfonts}
\usepackage{color}
\usepackage[colorlinks=true,linkcolor=magenta,urlcolor=magenta,citecolor=cyan,anchorcolor=blue]{hyperref}
\usepackage{floatrow}
\usepackage{multirow}
\begin{document}

\title{Steering-induced phase transition in measurement-only quantum circuits}
\author{Dongheng Qian}
\affiliation{State Key Laboratory of Surface Physics and Department of Physics, Fudan University, Shanghai 200433, China}
\author{Jing Wang}
\thanks{Corresponding author: wjingphys@fudan.edu.cn}
\affiliation{State Key Laboratory of Surface Physics and Department of Physics, Fudan University, Shanghai 200433, China}
\affiliation{Institute for Nanoelectronic Devices and Quantum Computing, Fudan University, Shanghai 200433, China}
\affiliation{Zhangjiang Fudan International Innovation Center, Fudan University, Shanghai 201210, China}

\begin{abstract}
Competing measurements alone can give rise to distinct phases characterized by entanglement entropy—such as the volume law phase, symmetry-breaking (SB) phase, and symmetry-protected topological (SPT) phase—that can only be discerned through quantum trajectories, making them challenging to observe experimentally. In another burgeoning area of research, recent studies have demonstrated that steering can give rise to additional phases within quantum circuits. In this work, we show that new phases can appear in measurement-only quantum circuits with steering. Unlike conventional steering methods that rely solely on local information, the steering scheme we introduce requires the structure of the circuit as an additional input. These steering induced phases are termed ``informative'' phases. They are distinguished by the intrinsic dimension of the bitstrings measured in each circuit run, making them substantially easier to detect in experimental setups. We explicitly show this phase transition by numerical simulation in three circuit models that were previously-studied: projective transverse field Ising model, lattice gauge-Higgs model and XZZX model. When the informative phase coincides with the SB phase, our steering mechanism effectively serves as a ``pre-selection'' routine, making the SB phase more experimentally accessible. Additionally, an intermediate phase may manifest, characterized by a discrepancy arises between the quantum information captured by entanglement entropy and the classical information conveyed by bitstrings. Our findings demonstrate that steering not only adds theoretical richness but also offers practical advantages in the study of measurement-only quantum circuits.
\end{abstract}

\date{\today}

\maketitle

\section{\label{intro}Introduction}
The interplay between unitary dynamics, measurements, and entanglement serves as a cornerstone in the study of fundamental quantum mechanics. As we transition into the Noisy Intermediate-Scale Quantum (NISQ) era~\cite{preskill2018}, characterized by the imminent realization of powerful quantum computers~\cite{arute2019,boixo2016}, the quantum circuit model emerges as an ideal platform for validating theoretical predictions through experiments. A case in point is the measurement-induced entanglement phase transition, a field to which substantial research has been devoted. These phase transitions have not only been simulated numerically~\cite{fisher2023,potter2022,li2018,li2019,gullans2020a,jian2020, bao2020,chan2019,jian2022,choi2020,bao2021a,bao2021b,garratt2023,lee2022,li2021a,li2023a,nahum2020,nahum2018,nahum2019,sharma2022,sang2021a,skinner2019,szyniszewski2019,vasseur2019,zabalo2020,zhou2019,zhou2020, sierant2022a, sierant2022b, sierant2023c, biella2021, weinstein2022,kelly2023} but also verified experimentally~\cite{koh2023,noel2021,czischek2021,roch2014}. Within this realm, measurement-only circuits stand out due to their lack of unitary evolution gates. Contrary to the prior belief that measurements primarily function to disentangle quantum states, recent findings reveal that non-commuting measurements alone can also lead to volume-law state~\cite{ippoliti2021}. Even when measurements hinder entanglement growth, conflicting measurements can still induce phase transitions between distinct area law phases, such as the symmetry-protected topological (SPT)~\cite{lavasani2021a,lavasani2022,lavasani2021b,sriram2022,zhu2023,kuno2023,klocke2022} and symmetry-breaking (SB) phases~\cite{lang2020,sang2021b,li2021b,roser2023,kuno2023,klocke2022}.

From the experimental perspective, the detection of these phases poses significant challenges. Typically, the relevant information is embedded in individual quantum trajectories, which makes the measurement of entanglement entropy complex and resource-intensive. Duplicating the same state multiple times is required; however, due to Born's rule, achieving the same trajectory again necessitates exponential resources—a problem often referred to as ``post-selection"~\cite{fisher2023,potter2022}. Various approaches have been suggested to mitigate these costs, such as the use of reference qubits~\cite{gullans2020b,dehghani2023} or employing cross entropy~\cite{li2023b}. Steering emerges as another promising strategy~\cite{buchhold2022}. In conventional measurement-based steering schemes, adaptive steering gates are applied based on the outcome of measurements, typically targeting a representative state ~\cite{roy2020, herasymenko2023, volya2023,morales2023}. With a proper steering scheme, the associated entanglement phase transition can be observed at the level of the density matrix. Furthermore, steering has been shown to give rise to entirely new phases~\cite{odea2022,ravindranath2023a,ravindranath2023b, sierant2023a, sierant2023b, piroli2023, iadecola2023}. In a unified framework, the first scenario could be interpreted  as one where steering is intentionally configured to make the boundary of the newly induced phase coincide with that of the entanglement phase. In this way, observing the phase transition induced by steering effectively serves as a measurement of the entanglement phase transition itself. It's worth noting that previous work mainly focuses on transitions from volum-law to area-law phases. The detection of phase transitions between different area-law phases, however, remains largely unexplored.

\begin{figure}[t]
\begin{center}
\includegraphics[width=3.4in, clip=true]{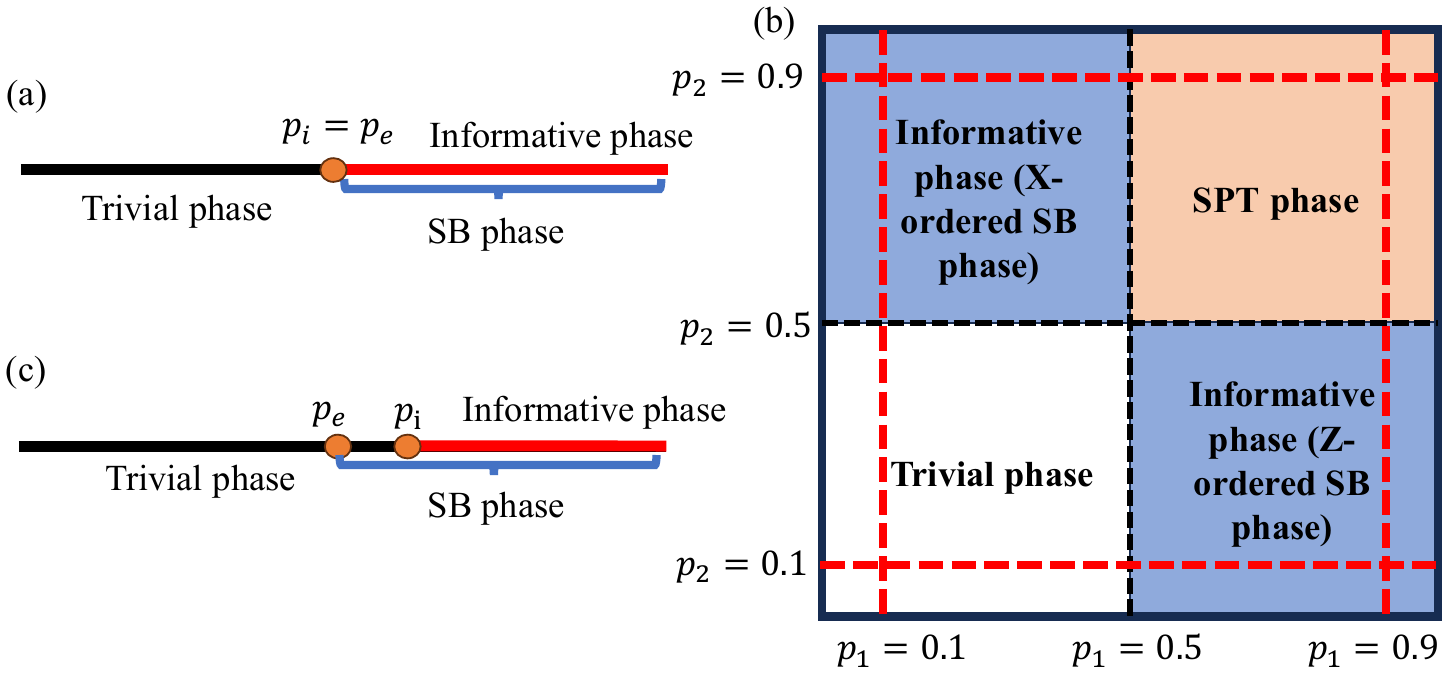}
\end{center}
\caption{Phase diagrams for different models considered in this work. (a) Phase diagram for the pTF-Ising model and XZZX model with only $X$ errors. $p_i$ means the critical point for informative phase transition while $p_e$ denotes the entanglement phase transition. The informative phase coincides with the SB phase. (b) Phase diagram for the lattice gauge-Higgs model. The red lines are the parameter scanning range in Sec.~\ref{gauge_higgs}. (c) Phase diagram for XZZX model with $Z$ errors only. $p_i$ and $p_e$ have the same meanings as in (a). There is an intermediate phase which is already symmetry-breaking and yet the bitstrings are not informative.}
\label{fig1}
\end{figure}

In this work, we explore phase transitions induced by steering in measurement-only quantum circuits. Specifically, we investigate the potential for steering to introduce an additional ``informative phase" within the SB phase~\footnote{The presence of symmetry-breaking in the system is characterized by the evolution of entanglement entropy. This phase is of non-equilibrium origin, allowing it to manifest even in one-dimensional systems.}. Our findings reveal that an appropriately designed steering scheme can indeed give rise to this new informative phase as shown in Fig.~\ref{fig1}, characterized by a significant reduction in the intrinsic dimension of bitstrings measured from the circuit's final states. In the informative phase, the intrinsic dimension is greatly reduced and this reduction capability reveals some information about the state, thus justifying the name ``informative". Our proposed steering scheme differs from conventional methods by requiring not just local measurement outcomes, but also the circuit's structural information as inputs, while past measurement outcomes remain irrelevant. The aim is to design a steering unitary gate that does not alter previously measured results, which we will illustrate with concrete examples later. The rationale for adopting a different steering scheme can be understood as follows. Earlier research primarily focused on steering within circuits but without competing measurements, and it often identified a ``dark state" towards which the circuit would eventually evolve into~\cite{odea2022,ravindranath2023a,ravindranath2023b,buchhold2022}. However, in measurement-only circuit with at least two competing measurement operations, there is no such dark state to steer toward, as any other measurement operation, regardless of how infrequent, would drive the state away.

We perform numerical simulations on three previously-studied measurement-only circuit models known for their entanglement phase transitions: the projective transverse field Ising (pTF-Ising) model~\cite{lang2020,sang2021b,li2021b,roser2023}, the lattice gauge-Higgs model~\cite{kuno2023}, and the XZZX model~\cite{klocke2022}. For the simulations, we begin with simple product states, and utilize stabilizer formalism for efficiency, given the absence of unitary gates in these circuits~\cite{aaronson2004,gottesman1998,gottesman1997}. Importantly, the critical points of the entanglement phase transitions in these models are unaffected by steering \cite{li2019}. Our numerical results allow us to locate the transition points of the informative phase and compare them with established entanglement phase transitions. In the pTF-Ising and lattice gauge-Higgs models, we discover that, with appropriate steering, the informative phase coincides with all SB phases. In the XZZX model, the coincidence of the informative phase with the SB phase persists when only $X$ errors are introduced. However, the introduction of $Z$ errors may result in an intermediate phase. This phase can be interpreted as a state in which  quantum information (reflected by non-trivial entanglement) is not captured by the measured bitstrings, which only reflect the state's classical information. This intermediate phase also appears when both $X$ and $Z$ errors are present. Remarkably, by adopting a stricter steering scheme—where every measurement in the circuit is subject to steering—the informative phase expands within the phase diagram while remaining being inside the SB phase. Based on these findings, we propose that the integration of steering into measurement-only quantum circuits is both experimentally beneficial and theoretically compelling. From an experimental standpoint, the informative phase transition is much more readily observable, thus providing a practical method for detecting SB phase boundaries when they coincide with informative ones. Notably, in the gauge-Higgs model, simply detecting the informative phase is sufficient to delineate all phase boundaries due to the absence of a direct phase transition between the trivial and SPT phases. Theoretically, future research could focus on identifying the conditions under which informative and SB phases coincide, extending these concepts to SPT phases or other complex phases, and developing an analytical framework to explain this novel phase transition.

The paper is organized as follows. In Section~\ref{setup}, we introduce our steering scheme implemented in this study. Detailed procedures for measuring bitstrings are provided, along with an explanation of how principal component analysis (PCA) is used to define the order parameter specific to the informative phase. In Section ~\ref{numsim}, we discuss the concrete models considered in this work, including pTF-Ising model, lattice gauge-Higgs model and XZZX model. Minor adjustments to our steering schemes provide numerical evidence that informative phases can occur within the SB phase. For the first two models, the boundary of the informative phase coincides with that of the SB phase, allowing for a clear distinction between all phases present within these models. In the case of the XZZX model, we present numerical evidence of an intermediate phase and discuss its physical implications. In Section~\ref{resource},  we detail a comparative analysis of the resource requirements for directly detecting the SB phase versus identifying the informative phase. Our findings suggest that when the informative phase overlaps with the SB phase, the phase transition can be easily observed experimentally, avoiding the complications associated with post-selection. Finally, we conclude our work with some further discussions in Section~\ref{discussion}. Some auxiliary materials are relegated to the Appendices.

\section{Setup}
\label{setup}

In this section, we introduce our settings, including the steering scheme and how we utilize PCA to define the order parameter for the informative phase. We begin by detailing the specific steering scheme employed in this study. Our discussion starts with a comprehensive overview of the general framework before delving into its particular implementation in pTF-Ising model as a concrete example. Subsequently, we outline the methodology for defining the order parameter of the informative phase, a process that involves two key steps: the acquisition of a bitstring followed by subsequent PCA of the data. The choice of operators for measuring the bitstring may vary depending on the model under consideration. 

\subsection{Steering scheme}
\label{steering}

In measurement-only quantum circuits, randomized measurements are selected from a predetermined set and applied at arbitrary positions. We only consider projective local measurements in this work. We assume every measurement can be described by a Pauli operator $\mathbb{M}$ and we take $\pm 1$ as the measurement outcomes. To be specific, the two projectors associated with $\mathbb{M}$ are $ (\mathbb{I}\pm\mathbb{M})/2$, with $\mathbb{I}$ is identity operator. Before a measurement $\mathbb{M}$ takes place (except for the first one), the state has a set of stabilizers denoted by $P$ that comes from our previous measurements. For example, this set could be:
\begin{equation}
P=\{\left(\mathbb{M}_1, 1 \right), \left(\mathbb{M}_2, -1 \right), \left(\mathbb{M}_3, -1 \right), \cdots\}.
\end{equation}
Here we write out the stabilizer and its sign separately to be clear. After $\mathbb{M}$ is measured, it's added to $P$ and stabilizers that anti-commute with $\mathbb{M}$ are excluded if there are any. We denote $-1$ as the outcome that needs to be steered and if that happens, the set of stabilizers now becomes $P'\cup \left(\mathbb{M}, -1 \right)$ where $P'$ only contains those stabilizers that commute with $\mathbb{M}$. The steering approach we employ in this study is designed to alter \emph{only the outcome of the specific measurement operator $\mathbb{M}$}, while leaving all other previously measured compatible outcomes unchanged. It's worth noticing that the main difference here is that steering scheme proposed in previous studies allow the steering unitary gate to change the results in $P'$~\cite{odea2022,ravindranath2023a,ravindranath2023b, iadecola2023}. To achieve this goal, we only need to find a Pauli operator $\mathbb{Q}$ that anti-commutes with $\mathbb{M}$ while commuting with the rest of stabilizers in $P'$ since 
\begin{equation}
\mathbb{MQ}\ket{\psi} = -\mathbb{QM}\ket{\psi} = \mathbb{Q}\ket{\psi},
\end{equation}
where $\mathbb{M}\ket{\psi} = -\ket{\psi}$. In this way, the sign in front of $\mathbb{M}$ becomes $+1$ while other signs stay the same. This search can be done by a classical computer with ease and a concrete example is given in the following. A few remarks are in order. First, the Pauli operator $\mathbb{Q}$ always exists since $P'\cup \left(\mathbb{M}, +1 \right)$ is also a valid set of stabilizers. It is usually not unique and any operator satisfying the condition above can be used as a proper steering gate. Second, the Pauli operator $\mathbb{Q}$ may have support on a large region. Nonetheless, it's still practical to realize in experiment since it's just a direct product of single qubit gates. Third, the search only uses the information of what kind of stabilizers are present, meaning that the steering scheme only needs the circuit's history structure as a further input. Although in the following numerical simulation, we focus on stabilizer circuit where the initial state is a stabilizer state, it is worth emphasizing that the steering scheme here requires no such constraint. The only difference is that whether $P$ can uniquely define a state or not. Meanwhile, since the outcome of the stabilizer measurement will not affect the entanglement phase transition in measurement-only circuit, the entanglement phase boundary remains the same under this steering scheme.

For pTF-Ising model~\cite{lang2020,sang2021b,li2021b,roser2023}, there are two kinds of competitive measurements in the circuit: $ZZ$-measurement and $X$-measurement. When the $ZZ$-measurement dominates, the system is in the SB phase. We do steering after every $ZZ$-measurement if the outcome is $-1$. To be concrete, we show how to find the appropriate steering gate in detail for this particular model. We set a bitstring $s = 000\cdots00$ at the beginning, and update it according to the following rule. If we measure $Z_{i}Z_{i+1}$, we can set $s_{i} = 1$ and if we measure $X_{i}$, we set $s_{i-1}=s_{i}=0$. When $Z_{i}Z_{i+1}$ measures out to be $-1$, we look at $s_{i-1}$ and $s_{i+1}$. If they are both $0$, we apply either $X_{i}$ or $X_{i+1}$ unitary gate at random. If only one of them is $0$, for example, $s_{i}$, then we apply $X_{i}$ unitary gate. If both of them are $1$, we choose either side at random, say, right. Then, we apply $X_{i+1},X_{i+2},\cdots,X_{m}$ unitary gates, until we meet $s_m=0$ or reach the boundary. Remark that there is nothing special in using the $X$ unitary gate, one could replace it with $Y$ unitary gate at wish. Without steering, the $ZZ$-measurement would bring sites into clusters of quasi-GHZ state. Take two qubits as an example, the state could be $\ket{00}+\ket{11}$, $\ket{00}-\ket{11}$, $\ket{01}+\ket{10}$, $\ket{01}-\ket{10}$. With steering, only the first two Ising-like states where all qubits are in the same state are now possible. Thus, the quasi-GHZ cluster now becomes Ising-like clusters and can be characterized by the order parameter to be defined in the following.

\subsection{Order parameter}
\label{order}

The order parameter to characterize the informative phase is defined as follows. After every run of circuit, we conduct a series of commuting measurements on the final state and get a bitstring $x_{i} = (x_{i}^{1}, x_{i}^{2}, \dots, x_{i}^{N})$ of length $N$. Noticing that since they commute with each other, we can simultaneously determine their values. With $M$ bitstrings $\{x_{i}\}$ at hand, we identify these as $M$ samples with $N$ features. We organize the dataset into a matrix $\mathbb{X}_{M\times N}$, and we conduct PCA on this dataset. PCA is one of the most commonly used unsupervised learning techniques and it has been widely used in many-body physics studies~\cite{pearson1901,rencher2002,hu2017,wang2016}. It was  also previously shown to be able to characterize the entanglement phase transition by making the whole state as the input~\cite{turkeshi2022}. Specifically, we find the eigenvector matrix $V$ that satisfies:
\begin{equation}
\mathbb{X}^{T}\mathbb{X}V = \lambda V
\end{equation}
which consists of $N'$ eigenvectors $\{v_{\alpha}\}$ corresponding to the largest $N'$ eigenvalues of covariance matrix $\mathbb{X}^{T}\mathbb{X}$. $N'$ is the reduced dimension number and $N' \ll N$.  We then conduct the dimensional reduction by $x'_{i} = x_{i}V$, which makes $x'$ only has $N'$ features now. The eigenvectors composed of $V$ are also called weighting vectors, they specify principle directions and the dataset has the largest variance after projecting onto these directions. Finally, the variance in the $n$-th principle direction is defined as 
\begin{equation}
\sigma_{n} = \text{var}\left(\{x_{i}^{'n}\}\right) = \sum_{i}\left(x_{i}^{'n} - \frac {\sum_{i}x_{i}^{'n}}{M}\right)^{2},
\end{equation}
where the biggest variance $\sigma_{1}$ is identified as the order parameter. We also use the second biggest variance $\sigma_{2}$ to locate the critical point in the following since it reflects the long-range fluctuation in the bitstring~\cite{hu2017} while $\sigma_{1}$ actually reflects the intrinsic dimension of the measured bitstrings from the information theory perspective~\cite{campadelli2015}. When the order parameter is high relative to the variance in other directions, it means that the data’s dimension can be effectively reduced. This ability of dimension reduction is the key information we utilize.

The last problem is how to choose the appropriate commuting measurements sets.  Thinking intuitively, the SB phase can be thought of as a $Z$-ordered phase where all qubits align with each other along the $z$-direction with steering. This reasoning naturally suggests that measuring the $z$-component could offer distinctive bitstring outcomes as the system transitions into the SB phase. More broadly speaking, one should focus on measurements along the axis that coincides with the direction of symmetry-breaking. The choice can be further validated by considering limiting cases, serving as a sanity check for the chosen measurement scheme. Take the pTF-Ising model as an example. With the aforementioned steering, the state is a simple superposition: $|\psi\rangle = |111\dots11\rangle \pm|000\dots00\rangle$  if only $ZZ$-measurements are present while being $\ket{+++\cdots++}$ if there are only $X$-measurements. Thus, we choose the set to be $Z$-measurement on every single qubit. In the first case, the bitstrings have only two possible configurations and thus the dataset is actually one-dimensional, leading to a high $\sigma_{1}$. In the second case, the measured bitstring is completely random and the variance in all directions would be almost the same, which makes it impossible to reduce its dimension. Knowing the limiting cases, the next step is to ask what would happen when both $ZZ$-measurements and $X$-measurements are present. We answer it with numerical simulation in the next section.

\section{Numerical Simulation}
\label{numsim}
\begin{figure}[t]
\begin{center}
\includegraphics[width=3.4in, clip=true]{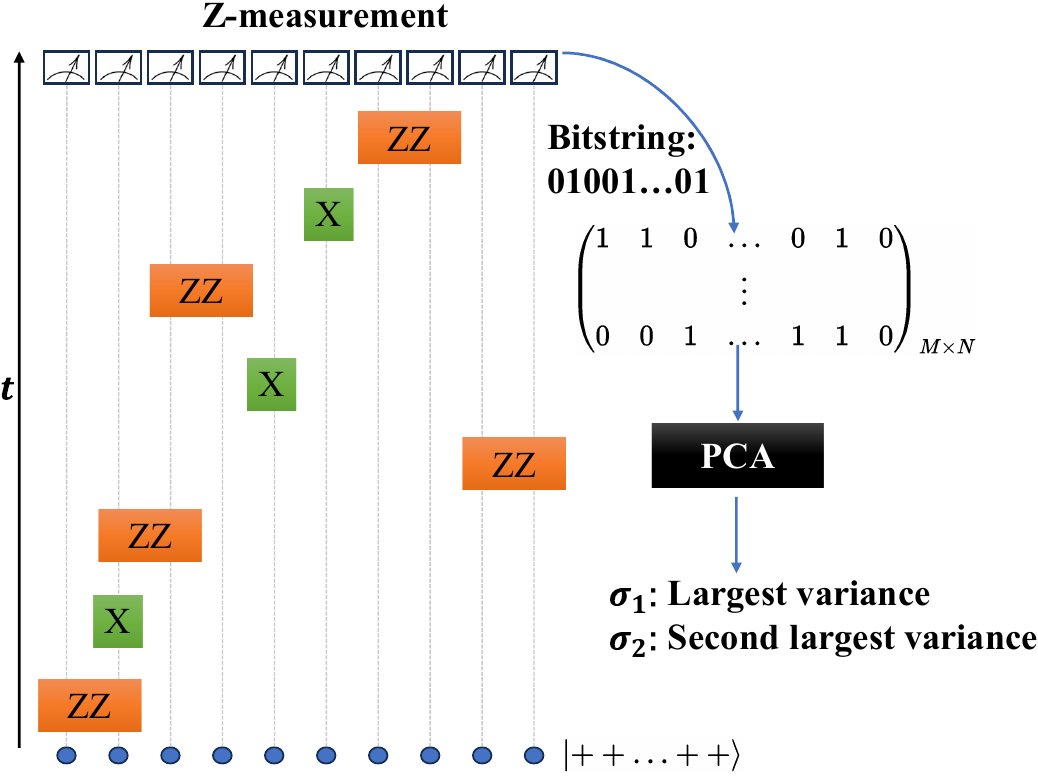}
\end{center}
\caption{A flowchart for the setup introduced in this work using pTF-Ising model as an example. The state is prepared initially in $x$-direction. Random $ZZ$-measurement and $X$-measurement are applied. At the end of the circuit, we measure every qubit in the $z$-direction and get a bitstring. By repeating this procedure many times, we get a dataset, and analyze it by PCA. Finally, $\sigma_1$ and $\sigma_2$ are read out.}
\label{fig2}
\end{figure}

After establishing the framework, we provide numerical evidence that steering can indeed induce an informative phase in various models by appropriately altering the set of commuting measurements performed at the end of the circuit. To illustrate that this phenomenon is quite general, we choose three different and previously-studied measurement-only circuit models: pTF-Ising model~\cite{lang2020,sang2021b,li2021b,roser2023}, lattice gauge-Higgs model ~\cite{kuno2023} and XZZX model~\cite{klocke2022}. The flowchart for numerical simulation is summarized and shown in Fig.~\ref{fig2}. To simulate large system sizes, we take the initial state to be stabilizer states in all the following models to take advantage of stabilizer formalism~\cite{aaronson2004,gottesman1998,gottesman1997}. For a circuit with length $L$, we evolve $O(L^2)$ time steps to get the final state and we conduct the simulation for system size up to $256$. Open boundary conditions are assumed throughout this work.

\subsection{projective transverse field Ising model}
\label{tfising_numerical}

\begin{figure*}
\begin{center}
\includegraphics[width=5.1in, clip=true]{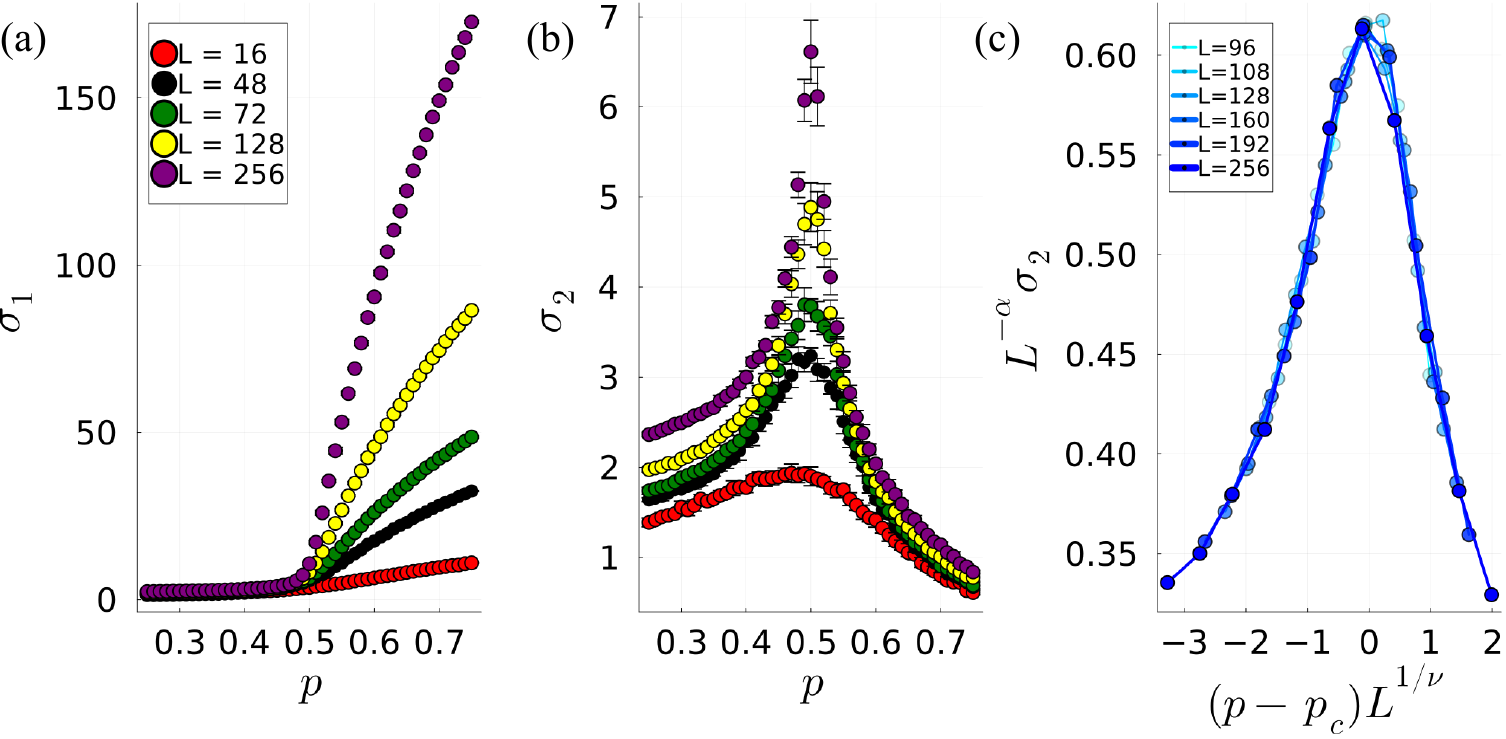}
\end{center}
\caption{Numerical results for pTF-Ising model. Every point is averaged over 10 runs and each run uses 1000 bitstrings as the input.  (a) $\sigma_1$ calculated by PCA. When $p<0.5$, it remains at a low value while abruptly rise at $p=0.5$, signifying a phase transition. Larger system size would lead to larger $\sigma_1$ as explained in the text. In the $p=1$ limit, $\sigma_1$ actually equals the system size $L$ since there are only two points in the data space. (b)  $\sigma_2$ calculated by PCA. The peak in $\sigma_2$ is at the phase transition point due to long-range fluctuation and is reminiscent of the susceptibility. (c) Data collapse for $\sigma_2$ used to locate the critical point and extract the critical exponent. Notice that we only use larger system size's data for data collapse to reduce finite-size effect.}
\label{fig3} 
\end{figure*}

In this model, we choose the initial state to be  $\ket{+++\cdots++}$.  At every time step, a $ZZ$-measurement is applied with probability $p$ at an arbitrary location. Otherwise, $X$-measurement is applied at an arbitrary position. If $ZZ$ measures to be $-1$, we then apply the appropriate steering gate. At the end of the circuit, we measure every qubit in the $z$-direction to get a bitstring. Every particular circuit structure is used only once, and the dataset is generated by 1000 such bitstrings. After that, we get the $\sigma_{1}$ and $\sigma_{2}$ as defined in Section~\ref{order}. To estimate the error, we repeat the above procedure for 10 times to get the average values and standard errors of $\sigma_{1}$ and $\sigma_{2}$ for every $p$ and $L$.  The result is shown in Fig.~\ref{fig3}(a). It can be seen clearly that $\sigma_{1}$ becomes significant at $p=0.5$. Larger system size would lead to larger $\sigma_{1}$ when $p>0.5$, since longer bitstrings $111\cdots111$ and  $000\cdots000$ are further away from each other after the projection. This is the first evidence of the presence of informative phase.

To locate the critical point more accurately, we look at $\sigma_{2}$ and collapse the data as shown in Fig.~\ref{fig3}(b,c). The detailed data collapse and error estimation procedure is described in Appendix~\ref{data_collapse}. We find that the critical point is at $p_c = 0.504 \pm 0.001$ and the critical exponent is $\nu = 1.36 \pm 0.08$, which coincides with the entanglement phase transition. While the entanglement phase transition can be mapped to bond percolation in two-dimensional square lattice,  it's worth mentioning that it's actually the same situation here. With steering, every $ZZ$-measurement is to cluster the neighboring qubits. The only difference is that by steering, the qubits in the cluster are now always in the same state, rather than constituting a general quasi-GHZ state. On the other hand, $X$-measurement singles out one qubit from the clusters. In the final bitstring, qubits that are connected by a cluster would have the same measurement outcome. 

\begin{figure}[b]
\begin{center}
\includegraphics[width=3.4in, clip=true]{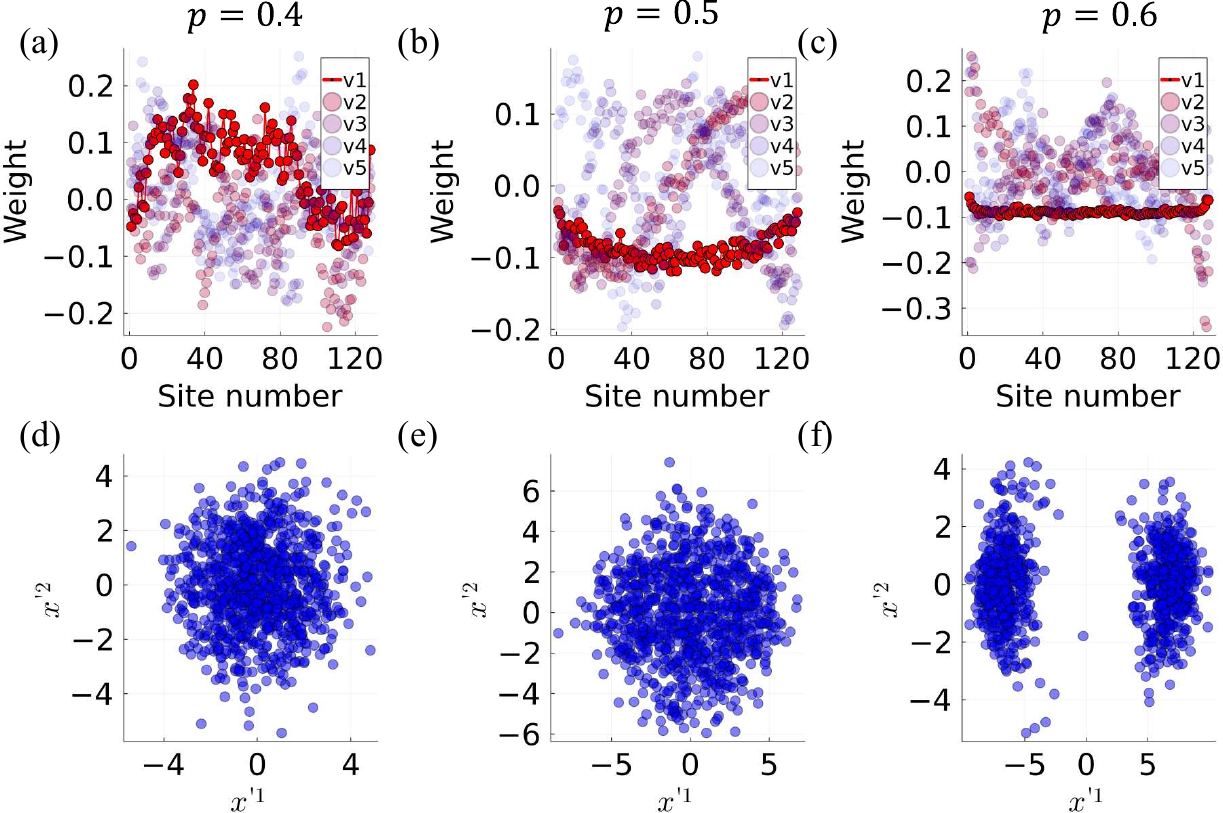}
\end{center}
\caption{PCA related results for pTF-Ising model. The system is $L=128$. (a)-(c) shows the first five principle weighting vectors and the biggest one is highlighted. One can see that the biggest weighting vector learns to discriminate bitstrings by adding the bitstring up as the system enters into informative phase. (d)-(f) is the projected data onto the first two principle directions for 1000 samples. One can clearly see that the data points are divided into two clusters in the informative phase (f).}
\label{fig4}
\end{figure}

Furthermore, the order parameter here can also be translated into percolation language. We explicitly show the eigenvectors $v_{\alpha}$ for the five largest directions in Fig.~\ref{fig4}(a)-\ref{fig4}(c).
As $p$ becomes greater than $0.5$, $v_{1}$ starts to be uniform. Recall that for every bitstring $x_{i}$, the projected first feature is:
\begin{equation}
x_{i}^{\prime 1} = v_{1} \cdot x_{i}.
\end{equation}
Thus, it means that the most important feature is actually just the total number of $1$s or $0$s in the bitstring when $p\geq0.5$ and fluctuations are comparably smaller. Putting it in another form, this feature is to check that \emph{whether there exists a large cluster of $1$s or $0$s in the bitstring}. This is exactly a signature of percolation transition~\cite{stauffer1994,cardy2001}. Finally, we directly project every bitstring to the first two directions in Fig.~\ref{fig4}(d) and Fig.~\ref{fig4}(e). One can visually see that in the informative phase, the data points are clustered around two points, which makes the dataset ``informative". This is reminiscent of the PCA result for classical Ising model. 

\subsection{lattice gauge-Higgs model}
\label{gauge_higgs}

\begin{figure*}
\begin{center}
\includegraphics[width=6.8in, clip=true]{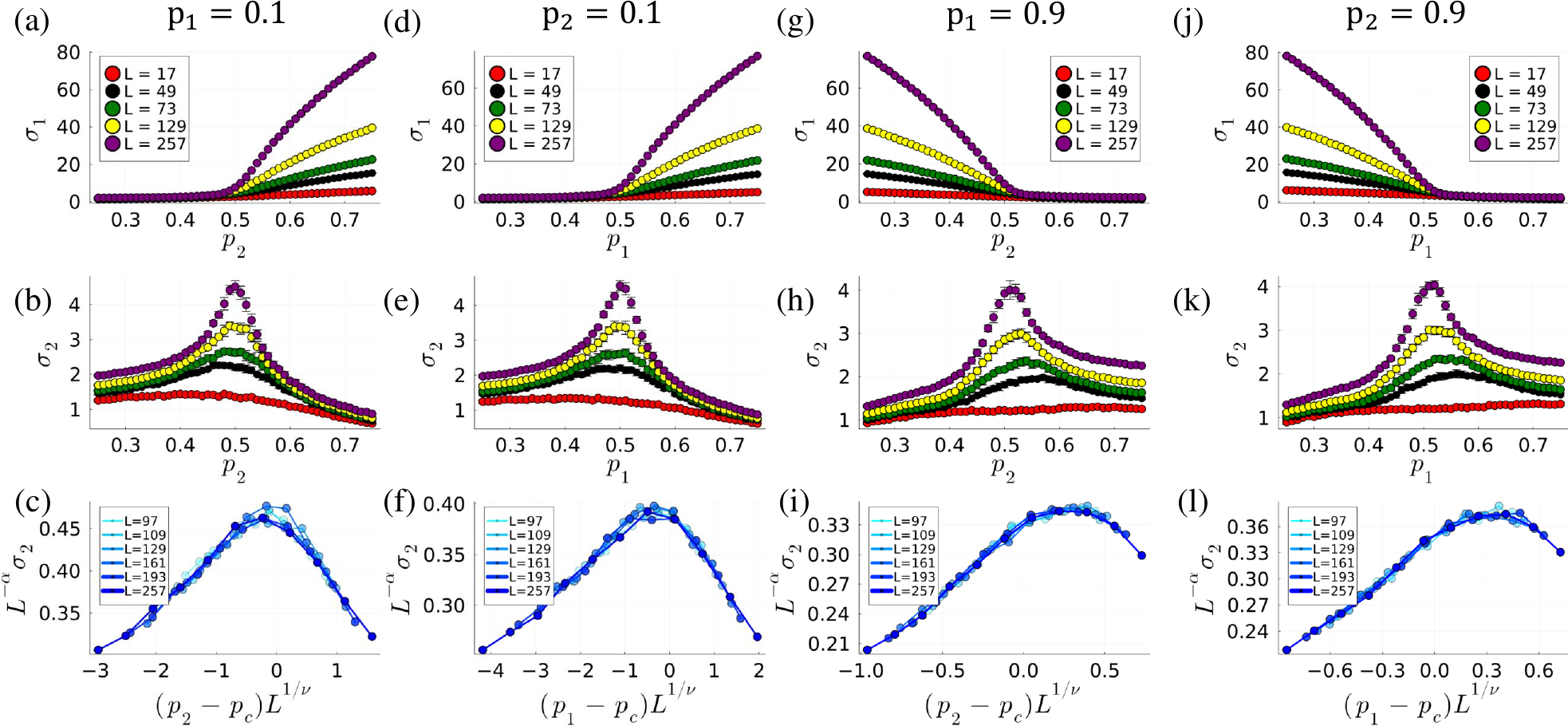}
\end{center}
\caption{Numerical results for lattice gauge-Higgs model. $p_1$ and $p_2$ are chosen according to the red dashed lines in Fig.~\ref{fig1}(b). Noticing that we only consider the case where there are odd number of qubits in the system for simplicity. First row is the result for $\sigma_1$, second row is the result for $\sigma_2$ and the third row is the data collapsing result. (a)-(c) $p_1=0.1$. When $p_2 > 0.5$, the system is in informative phase and coincides with the X-ordered SB phase. (d)-(f) $p_2=0.1$. When $p_1 > 0.5$, the system is in informative phase and coincides with the Z-ordered SB phase. (g)-(i) $p_1=0.9$. When $p_2 < 0.5$, the system is in informative phase and coincides with the Z-ordered SB phase. (j)-(l) $p_2=0.9$. When $p_1 < 0.5$, the system is in informative phase and coincides with the X-ordered SB phase. }
\label{fig5}
\end{figure*}

We next consider the lattice gauge-Higgs circuit model, which was first discussed in Ref.~\cite{kuno2023}. We leave aside the physical connection between this model and the original Hamiltonian and treat it merely as another measurement-only circuit with different measurement operations. A brief overview of the model's setup is provided here. For the sake of simplicity, we make the assumption that the system contains an odd number of qubits and we number them starting from 1. There are four kinds of measurement operations in this model: 
\begin{eqnarray}
\mathbb{M}_{1} &=& Z_{i}Z_{i+1}Z_{i+2} \quad (i~\text{mod}~2 = 0),
\nonumber
\\
\mathbb{M}_{2} &=& X_{i}X_{i+1}X_{i+2} \quad (i~\text{mod}~2 = 1),
\nonumber
\\
\mathbb{M}_{3} &=& X_{i} \quad (i~\text{mod}~2 = 0),
\nonumber
\\
\mathbb{M}_{4} &=& Z_{i} \quad (i~\text{mod}~2 = 1). 
\end{eqnarray}
When referring to these operations in the following, we assume they are acting on the correct sites implicitly. In this setup, the only two pairs of anti-commuting operators are $(\mathbb{M}_{1},\mathbb{M}_{3})$ and $(\mathbb{M}_{2},\mathbb{M}_{4})$, while other pairs commute with each other. During each time step, two measurements are taken from these paired sets. For the first measurement, $\mathbb{M}_1$ is applied with a probability of $p_1$ and $\mathbb{M}_3$ is applied otherwise. The location for the action is chosen randomly. For the second measurement, either $\mathbb{M}_2$ or $\mathbb{M}_4$ is  chosen in a similar fashion, controlled by a tuning probability $p_2$. There are both SB phases and SPT phase in this model, and the phase diagram is shown in Fig.~\ref{fig1}(b)~\footnote{In~Ref.~\cite{kuno2023}, X-ordered SB phase is called deconfinement phase, while Z-ordered SB phase is called spin-glass phase}.

\begin{table}[t]
\caption{We summarize different settings for different line scans in lattice gauge-Higgs model. The check mark means that the operator needs to be steered in that case. The rule to determine the setting is discussed in the main text. The critical point $p_c$ always means the unfixed parameter in the line scan. Both $p_c$ and $\nu$ are determined by the data collapse procedure. Notice that  $\nu$ for $p_1=0.9$ and $p_2=0.9$ are close to the critical exponent determined for the entanglement phase transition in Ref.~\cite{kuno2023}.}
\begin{center}\label{table1}
\renewcommand{\arraystretch}{1.7}
\begin{tabular*}{3.4in}
{@{\extracolsep{\fill}}ccccc}
\hline
\hline
& $p_1=0.1$ & $p_1=0.9$ & $p_2=0.1$ & $p_2=0.9$ \\
\hline 
$\mathrm{XXX}$ & $\checkmark$ & $\checkmark$ & $\checkmark$ & $\checkmark$ \\
$\mathrm{Z}$ & & $\checkmark$ & $\checkmark$ &  \\
$\mathrm{ZZZ}$ & $\checkmark$ & $\checkmark$ & $\checkmark$ & $\checkmark$ \\
$\mathrm{X}$ & $\checkmark$ &  & & $\checkmark$\\
Measure direction & $\hat{x}$ & $\hat{z}$ & $\hat{z}$ & $\hat{x}$ \\
$p_{\mathrm{c}}$ & $0.502(2)$ & $0.497(2)$ & $0.508(2)$ & $0.494(3)$ \\
$\nu$ & $1.45(8)$ &$1.96(9)$ &$1.35(9)$ &$2.0(1)$
\\
\hline
\hline
\end{tabular*}
\end{center}
\end{table}

Now we want to see whether informative phases are still present in this more complicated model. While all entanglement phase transitions in this model belong to the same universality class as those in the pTF-Ising model, the approach for identifying the informative phase necessitates slight modifications. The critical questions that arise are which measurements should be steered and what set of commuting measurement operators should be employed. Take the case where we fix  $p_2 = 0.1$ and tune $p_1$ as an example. In this scenario, we are looking at the competition between $\mathbb{M}_1$ and $\mathbb{M}_3$ while $\mathbb{M}_2$ and $\mathbb{M}_4$ are actually irrelevant. Thus, we steer $\mathbb{M}_1$  while leaving $\mathbb{M}_3$ unsteered similar to the pTF-Ising model. Meanwhile, we also steer the $\mathbb{M}_2$ and $\mathbb{M}_4$ to eliminate their influences. Moreover, since we are looking at the phase transition into Z-ordered SB phase, we choose to measure the $z$-component of the qubits. Other situations can also be determined in a similar way and the settings are summarized in Table~\ref{table1}.  In our numerical simulations, we concentrate on four specific line cuts within the parameter space, as indicated by the red lines in Fig.~\ref{fig1}(b). The results are shown in Fig.~\ref{fig5}. One can see that informative phase can indeed be found in this model.  We also locate the critical point and the critical exponent by data collapsing on $\sigma_2$. The results are shown in Table~\ref{table1} and the resultant phase diagram is shown in Fig.~\ref{fig1}(b). Remarkably, we find that the informative phase region again coincides with the SB phases. This indicates that the original entanglement phase transition is actually brought to a classical level by steering and only looking at the bitstring information. We argue that we can actually utilize this feature to make the SB phase more transparent to experiments. A comparative analysis is provided in Section~\ref{resource} to evaluate the resources required for observing the phase transition both with and without steering.

\subsection{XZZX model}
\label{xzzx}
Finally, we consider the XZZX measurement-only circuit model. This model was initially introduced and studied thoroughly in Ref.~\cite{klocke2022}. There are both SB phase and SPT phase in this model, and they could coexist in some parameter ranges. We mainly consider three different scenarios: cases with only $X$-errors, only $Z$-errors, and those where both types of errors are present. 

\begin{figure}[t]
\begin{center}
\includegraphics[width=3.4in, clip=true]{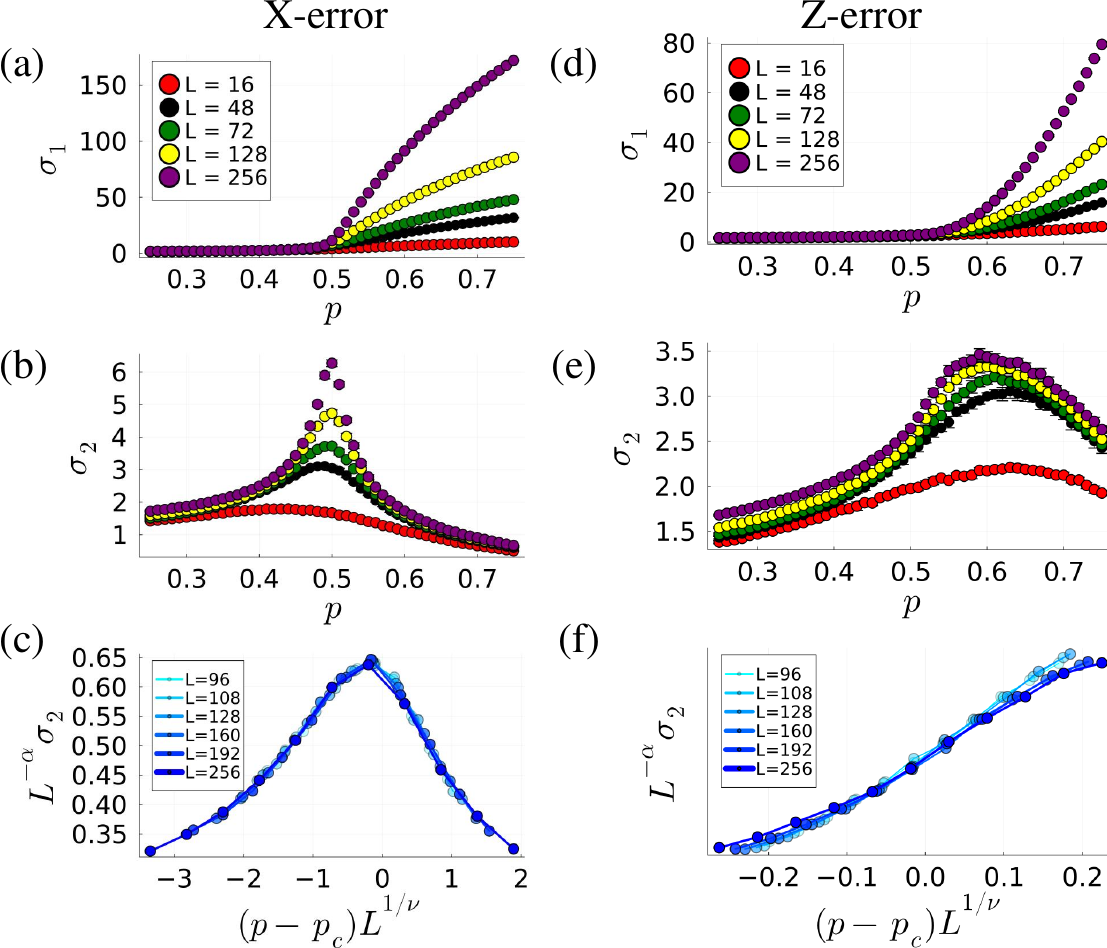}
\end{center}
\caption{Numerical results for XZZX model. (a)-(c) With $X$-error only. The informative phase again coincides with the SB phase. (d)-(f) With $Z$-error only. One can see that the shape of the curve is qualitatively different from the case with $X$-error. With data collapsing, we determine that $p_c=0.514\pm0.007$ and $\nu=3.5\pm0.7$. The universality class here remains unclear.}
\label{fig6}
\end{figure}

First, we consider the case where there are only one type of error in the circuit (meaning two competing measurements): 
\begin{eqnarray}
\mathbb{M}_{1} &=& X_{i}Z_{i+1}Z_{i+2}X_{i+3},
\nonumber
\\
\mathbb{M}_{2} &=& X_i,
\end{eqnarray}
and with probability $p$, $\mathbb{M}_1$ is measured at a random location. From the frustration graph point of view, this model is equivalent to the pTF-Ising model where the SB phase transition occurs at $p=0.5$. We choose to steer after every $\mathbb{M}_1$ measurement if the outcome is $-1$. The choice of commuting measurement set requires further consideration. Here, the symmetry-breaking ``direction'' is actually characterized by $XYX$. If one regards the $XZZX$ as $ZZ$ in the previous case, the proper ``$Z$" in this case now should be $XYX$ since $X_iZ_{i+1}Z_{i+2}X_{i+3} = X_iY_{i+1}X_{i+2} \cdot X_{i+1}Y_{i+2}X_{i+3}$. Thus, the proper set of commuting measurements at the end of each circuit run should now be $\{X_{i}Y_{i+1}X_{i+2}\}$. Since we are considering open boundary conditions, the measured bitstring would have the length $N=L-2$ where $L$ is the total qubit number and $x_i^{n}$ is now the measurement outcome of $ X_iY_{i+1}X_{i+2}$. Under this setting, the result is shown in Fig.~\ref{fig6}(a-c).
The critical point is found to be at $p_c=0.506 \pm 0.001$ and the critical exponent is $\nu = 1.33 \pm 0.08$. One can see that the informative phase  coincides with the SB phase as it does in previous models. However, the situation changes if we consider the case where $\mathbb{M}_2 = Z_i$. The SB phase transition still occurs at $p=0.5$, and there is also SPT phase transition at the same point. For the informative phase, as is shown in Fig.~\ref{fig6}(d-f), the phase transition point is now  at $p_c = 0.514 \pm 0.007$. Thus, we suspect that an intermediate phase may appear and the phase diagram is shown in Fig.~\ref{fig1}(c). In the intermediate phase, although the system exhibits a symmetry-breaking nature, as evidenced by specific entanglement measures, this characteristics does not readily translate to the bitstrings. In other words, the classical information carried by these bitstrings is insufficient to reveal the underlying SB phase within this intermediate zone. It is worth noticing that the circuit with $Z$-error does not have a percolation picture when looking at the measured bitstrings. The impact of a $Z$-measurement is considerably more intricate than that of an $X$-measurement. Unlike an $X$-measurement, which typically isolates a single qubit from an $XYX$-cluster, a $Z$-measurement has the capacity to reconfigure the states of adjacent qubits, potentially forming an entirely new cluster. A more detailed example is given in Appendix~\ref{z_error}. Thus, there's no constraint that the informative phase transition should coincide with the SB phase transition in this scenario.

\begin{figure}[t]
\begin{center}
\includegraphics[width=3.4in, clip=true]{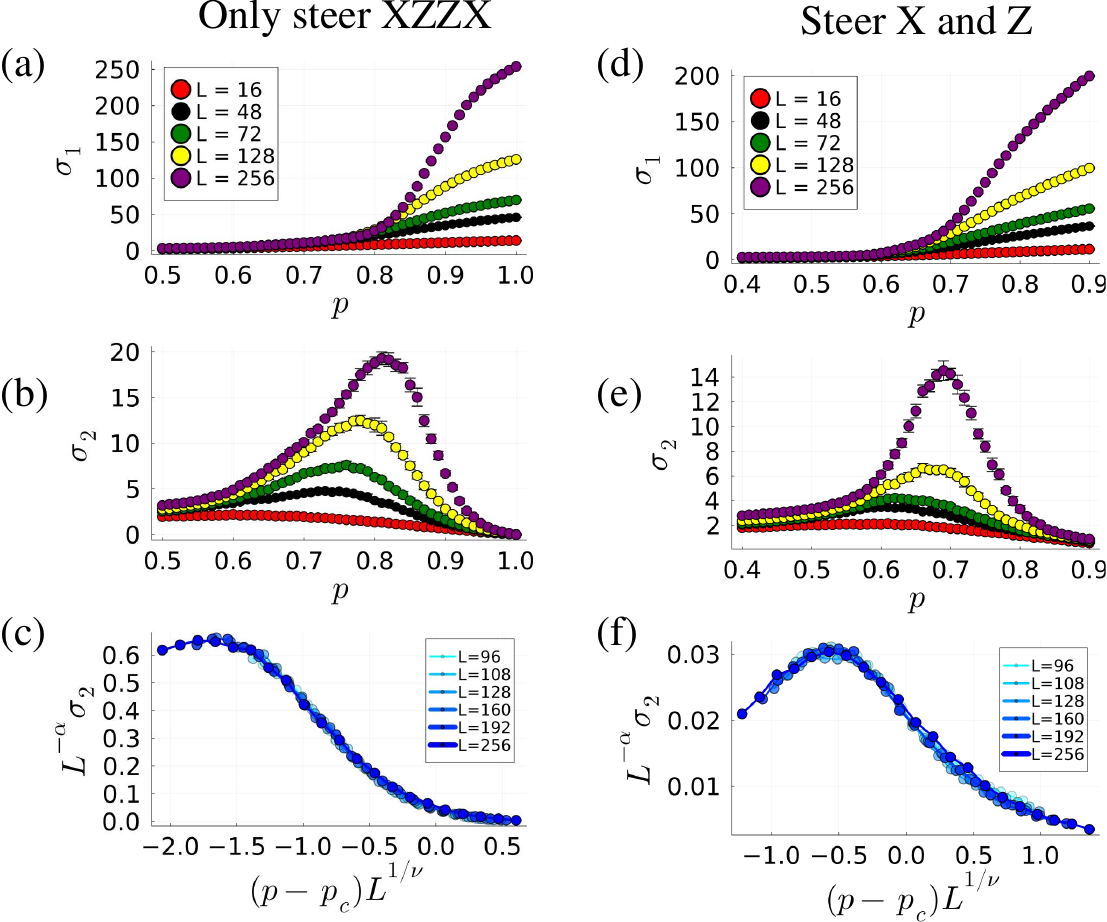}
\end{center}
\caption{Numerical results for XZZX model with both $X$-error and $Z$-error at different steering strength. (a)-(c) Steering gate only applies when $XZZX$ measures to be $-1$. We determine $p_c=0.945\pm0.003$ and $\nu=2.14\pm0.07$ which suggests the presence of an intermediate phase. (d)-(f) Steering gate is applied after every measurement if the measurement outcome is $-1$. We have $p_c=0.736\pm0.002$ and $\nu=2.2\pm0.1$ now. It clearly shows that the region of informative phase grows.}
\label{fig7}
\end{figure}

Now we consider a more complicated situation where there are both $X$-measurement and $Z$-measurement in the circuit. For simplicity, we assume that the probabilities for them to occur are equal and the circuit structure is now captured by:
\begin{eqnarray}
\mathbb{M}_{1} &=& X_{i}Z_{i+1}Z_{i+2}X_{i+3},~\text{with probability}~p
\nonumber
\\
\mathbb{M}_{2} &=& X_i,~\text{with probability}~\frac {1-p}{2}
\nonumber
\\
\mathbb{M}_{3} &=& Z_i,~\text{with probability}~\frac {1-p}{2}.
\end{eqnarray}
In this case, SB phase transition was shown to occur at $p_c \approx 0.56$ and the SPT phase is absent as long as $p\ne 1$. We first choose to only steer on $\mathbb{M}_1$ as was previously done. The result is shown in Fig.~\ref{fig7}(a-c). The critical point for informative phase transition here is determined as  $p_c=0.945 \pm 0.003$ by data collapsing. Thus, an intermediate phase apparently appears. Traditionally, steering strength is modulated by a probability term which governs whether a steering gate is applied after an undesired measurement outcome. Increasing this probability generally leads to a more expansive region for the steering-induced phase within the phase diagram~\cite{ravindranath2023a, odea2022}. In our models, we find that when the steering probability is anything other than one, no informative phases emerge (data not shown). Alternatively, we investigate the effect of modifying steering strength by selectively applying steering to a greater or lesser number of measurement operators. As an example, we choose to steer on all the measurements in the circuit and the result is shown in Fig.~\ref{fig7}(b). The critical point is now greatly shifted to be at $p_c=0.736\pm0.002$. As expected, the informative phase indeed grows in region with a stronger steering strength while still staying inside the SB phase. Our findings suggest that enlarging the set of measurements subjected to steering results in a broader informative phase region.

\section{Comparison of resources}
\label{resource}

Having demonstrated that the informative phase boundary consistently occurs within the SB phase in three distinct models, we are now poised to think about the physical significance of the coincidence of phase boundaries of SB phase and informative phase. It was previously pointed out that one could make the entanglement phase transition coincide with another easy-to-measure classical ordering transition by steering the system into a unique representative wave function, which is named as ``pre-selection''~\cite{buchhold2022}. The situation here is similar, but with a major difference that the representative wave function is not unique. The competition was usually between unitary gates and a single measurement operation in previous studies, while here we are considering competition between different measurements. Take $ZZ$-measurement and $X$-measurement as an example. Although a reasonable target representative state is $\ket{\psi}=\ket{111\cdots11}\pm\ket{000\cdots00}$, the $X$-measurement would drive the state away from it  as long as the probability of an $X$-measurement occurring is non-zero. Thus, the steering can be seen as a generalization of ``pre-selection'' to measurement-only circuit when the informative phase coincides with the SB phase. It is then natural to ask whether detecting the informative phase requires fewer experimental resources than directly detecting the SB phase. We now compare them in detail.

\begin{table}[t]
\caption{Comparison for the cost of resources to observe SB phase transition and the informative phase transition. Costs in the same column are multiplied to get the final cost. Notice that for informative phase, we ignore the cost of finding the proper steering gate to act which is relatively small. We also overestimate the cost of applying the steering gate by assuming that every gate has the support on the whole system, which is usually not the case.}
\begin{center}\label{table2}
\renewcommand{\arraystretch}{1.7}
\begin{tabular*}{3.4in}
{@{\extracolsep{\fill}}ccc}
\hline
\hline
& SB & Informative \\
\hline Post-selection & $2^{L^2}$ & $p L^3$ single qubit gate \\
Trajectory & $\sim1000$ & \multirow{2}*{ $\sim 1000$}  \\
Circuit & $>100$ &  ~ 
\\
\hline
\hline
\end{tabular*}
\end{center}
\end{table}

The cost of directly measuring the SB phase transition is threefold. First, one needs to get the same trajectory by post-selection, which is exponential in the system size. Second, the same trajectory needs to be prepared many times to determine its property, such as expectation values or entanglement entropy. Finally, one needs to average over many different random circuits. The total cost is the product of the three. In contrast, for the informative phase, we significantly reduce these complexities. We get rid of the need to post-select by applying a series of steering gates.  Moreover, we only need to run a single circuit for once. In the numerical simulation above, the $1000$ bitstrings come from different random circuits. This comparison is summarized in Table~\ref{table2}. Thus, we argue that the resource needed to observe the informative phase transition is  much less than the SB phase transition. A notable consequence is observed in the context of the lattice gauge-Higgs model. The absence of a direct boundary between the trivial phase and the SPT phase, combined with the complete overlap of the informative phase with the SB phases, offers a unique advantage for experimental measurements. Specifically, all the phase transitions in this model become experimentally accessible through the use of steering, followed by PCA on the measured bitstrings.

\section{Summary and Discussion}
\label{discussion}

In this study, we integrate steering mechanisms into measurement-only quantum circuits to explore their phase behavior. We discover an informative phase, defined via an easily measurable order parameter related to the intrinsic dimensions of bitstrings. Utilizing a specialized steering scheme that requires non-local information, we demonstrate the emergence of these informative phases within existing SB phases. Our findings are substantiated through numerical simulations across three distinct models. When the boundaries of these phases overlap, our approach offers a viable method for experimentally detecting entanglement phase transitions, serving as a form of pre-selection without the need for a single ``dark state''. Additionally, we identify the potential for an intermediate phase that exhibit symmetry-breaking characteristics in terms of entanglement, yet their bitstrings remain non-informative. For this case, it is interesting to ask whether there could be some phases that coincide with the SB phase by applying another steering scheme and looking at other easy-to-measure order parameter. A future direction for study involves exploring how we can classify models based on the possibility of revealing the SB phase transition in an experimentally feasible way.

The deep reason that the informative phase transition can be observed with ease in experiment can be summarized as follows. By steering, we narrow down the possible final outcomes of the circuit to a certain subset of all possible ones. Thus, the final state should be described by a mixed state rather than a single pure state. After that, we look at whether the state has a large cluster where all qubits are in the same state by PCA. While the particular cluster size and location is different across all the pure states comprising the mixed state, \emph{the information whether or not there exists a large cluster is shared by all of them.} This is the key point why we don’t need post-selection to observe such transition. Meanwhile, this also suggests that the informative phase coming from this recipe should always lie inside the SB phase, since there couldn’t be a large cluster if there is no symmetry-breaking. The next question naturally arises: could there also be steering-induced phases within SPT phases? This presents a challenge since, in SPT phases, information is globally encoded and shielded from local measurements. Consequently, gaining insights from bitstrings obtained through local commutative measurements may prove difficult. One avenue for future research might involve non-local commutative measurements. Alternatively, a more sophisticated approach could seek to identify high-level features akin to the large clusters found in SB phases, but specific to SPT phases. We leave this subject to future work.

To define the information carried by bitstrings, we use PCA and think of the ability to reduce the data’s dimension as the useful information contained in them. It is reminiscent of discovering both quantum and classical phase transitions by various unsupervised machine learning methods~\cite{carleo2019,carrasquilla2020,mehta2019}. There, the bitstrings are generated either in experiments or by Monte Carlo sampling. It is interesting to ask whether using other more advanced techniques would squeeze more information from the bitstring such as diffusion map~\cite{lidiak2020,scheurer2019}, two-NN~\cite{facco2017,dalmonte2021,rodriguez2021,vitale2023}. A notable difference is that these methods usually work when sampling over a partition function, where the ground state always has the highest probability to occur. In our case, however, no such ground state appears. For example, in the pTF-Ising model, the state where all qubits  point in the same direction is always not the most probable state to occur unless $p=1$. We conjecture that this may prevent other methods from working better than PCA.

\begin{acknowledgments}
We acknowledge helpful discussions with Xiao-Liang Qi, Biao Lian and Yang Qi. This work is supported by the National Key Research Program of China under Grant No.~2019YFA0308404, the Innovation Program for Quantum Science and Technology through Grant No.~2021ZD0302600, the Natural Science Foundation of China through Grant No.~12174066, the Science and Technology Commission of Shanghai Municipality under Grant No.~20JC1415900 and No.~23JC1400600, and Shanghai Municipal Science and Technology Major Project under Grant No.~2019SHZDZX01.
\end{acknowledgments}

\begin{appendix}

\section{Data Collapse}
\label{data_collapse}

\begin{figure}[t]
\begin{center}
\includegraphics[width=2.4in, clip=true]{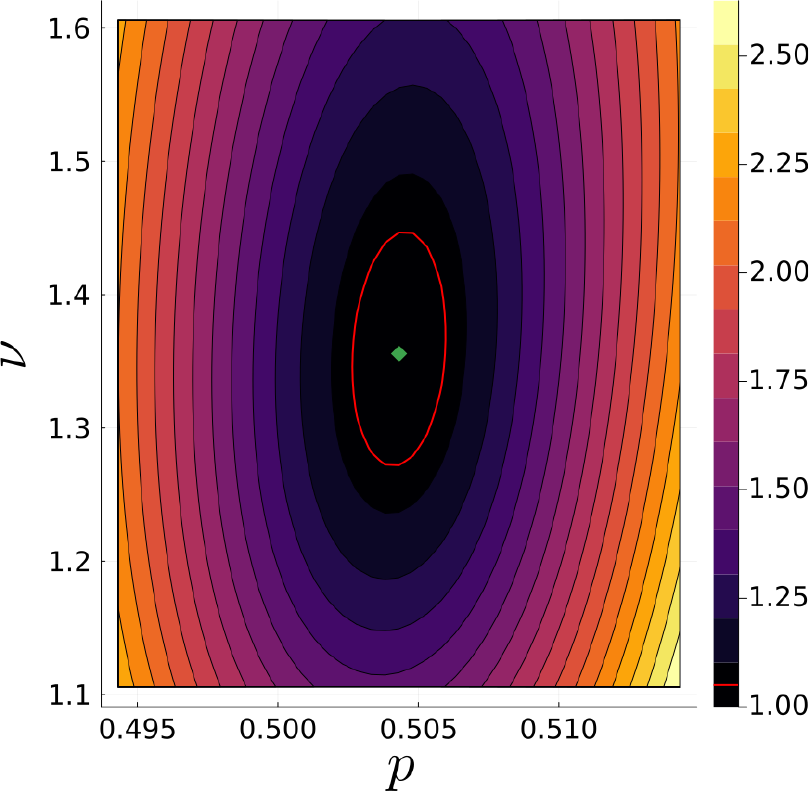}
\end{center}
\caption{Determine the uncertainty in $p_c$ and $\nu$. The residues are rescaled by dividing the $\epsilon^{\text{min}}$. Red circle is where the residue equals $1.05\epsilon^{\text{min}}$ and the green diamond is where the minimal point is. The uncertainty can be approximated to be $p_c=0.504\pm 0.001$ and $\nu=1.36 \pm 0.08$.}
\label{fig8}
\end{figure}

For $\sigma_2$, we hope to fit the curve according to a proper scaling hypothesis that $\sigma_2 = L^{\alpha}f((p-p_c)L^{1/\nu})$. Here $\alpha$ characterizes the degree of divergence while $\nu$ is the usual critical exponent. The data collapse procedure goes as follows. For a given combination of $p_c$, $\nu$ and $\alpha$, we can rescale a particular data point $(p, L, \sigma_2)$ to be: 
\begin{equation}
    x = (p-p_c)L^{1/\nu},\ \ y=\sigma_2L^{-\alpha}.
\end{equation}
After rescaling all the data points, we fit the rescaled data with a $12$-th order polynomial and we get the residue for the best fit. The residue $\epsilon(p_c, \nu, \alpha)$ is then defined as the target function. By applying the Nelder-Mead algorithm, we find the minimal point $(p_c^{\text{min}}, \nu^{\text{min}}, \alpha^{\text{min}})$ and the minimal value $\epsilon^{\text{min}}$. 

To estimate the uncertainty in $p_c$ and $\nu$, we fix $\alpha=\alpha^{\text{min}}$ and draw out $\epsilon=\epsilon(p_c, \nu, \alpha^{\text{min}})$. We take the threshold to be $1.05\epsilon^{\text{min}}$ to determine the uncertainty. Fig.~\ref{fig8} shows an example for estimating the error for pTF-Ising model.

\section{Effect of $Z$-error}
\label{z_error}

We give an illustrative example to see that the effect of $Z$-error on the bitstring is different from the case for $X$-error. Consider a system with 8 qubits. Starting from the state where every qubit is in $\ket{1}$, we first measure $XZZX$ at all possible positions. If we measure the bitstring by $\{XYX\}$ now, the result would be all $1$s or $0$s since the whole system is in one $XYX$-cluster. Now, if we further measure $Z$ on the fifth qubit and we assume that the outcome is 1, the state can be represented by stabilizers: 
\begin{eqnarray}
\{&&X_1Z_2Z_3X_4,X_3Z_4Z_5X_6,X_4Z_5Z_6X_7,X_2Z_3Z_4Z_6Z_7X_8,
\nonumber
\\
&&Z_3Z_6, Z_1Z_3Z_7,Z_2Z_5Z_8, Z_5\}.
\end{eqnarray}
Then, we measure the bitstring by measuring $\{XYX\}$ and denote the result of $X_iY_{i+1}Z_{i+2}$ as $x_i$. The possible outcomes are ${111111, 110001, 001110, 000000}$. One would find that $\{{x_1,x_2,x_6}\}$ are in the same cluster while $\{{x_3, x_4, x_5}\}$ are in another cluster. Thus, $Z$-error creates a new cluster instead of simply driving one qubit out of the $XYX$ cluster in this simplest case.  Moreover, the situation becomes more complicated if one again measures $Z$ on, for example, the sixth qubit and it eventually lose a simple picture to describe its effect. 

\end{appendix}

%

\end{document}